\documentclass[aps,prb,nobibnotes,twocolumn,superscriptaddress,bibliography]{revtex4-2}

\pdfoutput=1

\usepackage{chemformula} % Formula subscripts using \ch{}
\usepackage[T1]{fontenc} % Use modern font encodings

\usepackage{amsfonts}
\usepackage{mathrsfs}
\usepackage{amsmath}% needed for subequations
\usepackage{color}
\usepackage{graphicx}
\graphicspath{ {./pictures/v1/} }
\usepackage{bm}% bold maths
\usepackage{amssymb}
\usepackage{xspace}
\usepackage{epstopdf}
\usepackage{dcolumn}% Align table columns on decimal point
\usepackage{longtable}
\usepackage{multirow}
\usepackage{float}
\usepackage{comment}
\usepackage{lipsum}
\usepackage{tabularx}
\usepackage{booktabs}
\def\ie{{\it i.e.},\ }

\def\ea{{\it et al.}}
\usepackage{physics}
\usepackage[colorlinks=true, letterpaper=true, pdfstartview=FitV, linkcolor=blue, citecolor=blue, urlcolor=blue]{hyperref}

\newcommand{\Rom}[1]{\uppercase\expandafter{\romannumeral#1}}

\makeatother

\begin{document}
\title
{The electrical transport of intrinsic two-dimensional ferroelectric metal \ch{PtBi2}}

\author{Dan Li}
\affiliation{Research Center for Quantum Physics and Technologies, Inner Mongolia University, Hohhot 010021, China}
\affiliation{Inner Mongolia Key Laboratory of Microscale Physics and Atomic Manufacturing, Inner Mongolia University, Hohhot 010021, China}
\affiliation{School of Physical Science and Technology, Inner Mongolia University, Hohhot 010021, China}

\author{Liu Yang}
\email{yangliu1@ctgu.edu.cn}
\affiliation
{Department of Physics, Hubei Engineering Research Center of Weak Magnetic-field Detection, China Three Gorges University, Yichang, 443002, China}

\author{Lei Li}
\email{lilei1993@imu.edu.cn}
\affiliation{Research Center for Quantum Physics and Technologies, Inner Mongolia University, Hohhot 010021, China}
\affiliation{Inner Mongolia Key Laboratory of Microscale Physics and Atomic Manufacturing, Inner Mongolia University, Hohhot 010021, China}
\affiliation{School of Physical Science and Technology, Inner Mongolia University, Hohhot 010021, China}

\begin{abstract}
	
Breaking the conventional stereotype that ferroelectrics are necessarily insulating, two-dimensional (2D) ferroelectric metals combine seemingly incompatible switchable electric polarization and metallic conductivity, providing a fertile ground for the discovery of novel electrical transport phenomena and the development of innovative electronic devices. 
Using the semiclassical Boltzmann equation and first-principles calculations, we systematically investigate the linear and nonlinear transport responses of the intrinsic 2D ferroelectric metal \ch{PtBi2} to an applied electric field. 
Our \textit{ab initio} molecular dynamics simulations reveal that it possesses a high Curie temperature reaching $800~\text{K}$. 
We propose that the crystal structure of its high-temperature paraelectric phase can be explicitly distinguished through simple measurements of the in-plane electrical conductivity. 
Quantitative calculations of the Edelstein effect and the intrinsic spin Hall effect demonstrate a sizable charge-to-spin conversion efficiency, highlighting its potential in spintronics. 
We also find that a Berry curvature dipole-induced nonlinear Hall effect emerges in uniaxially strained \ch{PtBi2}. 
Furthermore, we highlight the unique advantages of 2D ferroelectric metals in gate-controlled transport applications. Based on the domain wall scattering mechanism, we conceptually design a novel ferroelectric metal field-effect transistor (FEM-FET) capable of nonvolatile switching between high-resistance and low-resistance states under a gate voltage. 
Our work not only unveils the rich transport physics in 2D ferroelectric metals but also provides valuable insights into the design of next-generation nonvolatile memory and spintronic devices.
	
\end{abstract}

\maketitle

\section{Introduction}
When scaled down to the atomically thin limit, materials often exhibit physical properties that are fundamentally distinct from those of their bulk counterparts. 
A prototypical example is graphite: when exfoliated down to a single atomic layer, the resulting graphene \cite{geim2007rise,castro2009electronic,geim2009graphene} exhibits massless Dirac fermions \cite{novoselov2005two,novoselov2004electric}, record thermal conductivity \cite{balandin2008superior}, and exceptional stiffness \cite{lee2008measurement}.
Another remarkable case is molybdenum disulfide ($\ch{MoS2}$), which undergoes a transition from an indirect to a direct bandgap semiconductor at the monolayer limit \cite{PhysRevLett.105.136805}, where the inherently broken inversion symmetry gives rise to novel spin-valley physics \cite{schaibley2016valleytronics,zeng2012valley,xu2014spin}.
This dimensional effect is an important reason why two-dimensional (2D) materials \cite{mounet2018two,miro2014atlas} have emerged as a prominent research hotspot in modern condensed matter physics.

Trigonal \ch{PtBi2} ($t$-\ch{PtBi2}) \cite{10.1063/5.0272618} is a van der Waals layered material that may serve as another excellent platform for demonstrating  such dimensional effects.
In its bulk form, $t$-\ch{PtBi2} is a topological semimetal hosting triply degenerate points \cite{gao2018possible} and Weyl fermions \cite{veyrat2023berezinskii}, and its superconductivity also attracts extensive experimental attentions \cite{changdar2025topological,moreno2025robust,schimmel2024surface,10.1063/10.0014014}. 
Intriguingly, when thinned down to a single monolayer, \ch{PtBi2} was predicted to retain its in-plane metallic conductivity while developing a spontaneous out-of-plane polarization induced by its noncentrosymmetric structure \cite{PhysRevLett.133.186801}. 
This unique combination renders monolayer \ch{PtBi2} a long-sought intrinsic 2D ferroelectric (FE) metal \cite{fei2018ferroelectric,24j9-8g8v,liu2026ferroelectricity}. 
Generally, the strong screening effect of itinerant electrons in three-dimensional (3D) metals quenches any macroscopic polarization, meaning that conventional 3D ferroelectrics are almost exclusively insulators \cite{lines2001principles}. 
Because of their insulating nature, the electrical transport properties of traditional ferroelectrics are inherently limited and rarely studied. 
Although the coexistence of ferroelectricity and metallicity can be realized in 2D materials, such intrinsic FE metals have been historically scarce. 
Consequently, the interplay between FE polarization and metallic transport remains a significant blank space in condensed matter physics, leaving the electrical transport of 2D FE metals largely unexplored and providing a fertile ground for discovering novel phenomena.

In this paper, based on the semiclassical Boltzmann equation, we systematically investigate the linear and nonlinear transport responses of the 2D FE metal \ch{PtBi2} to an applied electric field. 
Specifically, we examine the linear responses of the electric current, spin magnetization, and spin current, which correspond to the electrical conductivity, the Edelstein effect, and the spin Hall effect, respectively.
Regarding nonlinear transport phenomena, we focus on the nonlinear Hall effect (NLHE).
Symmetry analysis reveals that the Edelstein effect and the NLHE are odd under spatial inversion symmetry $\mathcal{P}$, intimately coupling them to the spontaneous FE polarization. 
Using first-principles electronic band structures and effective Hamiltonian, we quantitatively evaluate these response coefficients.
Building on these physical insights, we further propose a conceptual design for a gate-controlled electrical transport device that can switch between high-resistance and low-resistance states by modulating domain wall scattering in 2D FE metals.

The remainder of this paper is organized as follows. In Sec.~\ref{section:method}, we introduce the computational methods and the details of our first-principles calculations. 
In Sec.~\ref{section:curie}, we evaluate the Curie transition temperature of the 2D FE metal \ch{PtBi2} using \textit{ab initio} molecular dynamics. 
In Secs.~\ref{section:cond} through \ref{section:NLHE}, we employ the semiclassical Boltzmann transport theory to derive the formulas for the conductivity, the Edelstein effect, the spin Hall effect, and the NLHE, and we quantitatively calculate their coefficients as functions of the Fermi level.
In Sec.~\ref{section:FET}, we propose a novel gate-controlled electrical transport device, termed a ferroelectric metal field-effect transistor (FEM-FET).
Finally, Sec.~\ref{section:conclusion} contains a brief summary and discussion.

\section{Computation Method}\label{section:method}
Density functional theory (DFT) calculations involved in this work were implemented in the Vienna \textit{ab initio} Simulation Package (VASP 5.4.4) code \cite{vasp1,vasp2}. 
The exchange and correlation interactions were described by Perdew-Burke-Ernzerhof (PBE) \cite{GGA} functional based on the generalized gradient approximation (GGA).
Projector-augmented wave (PAW) method  \cite{PAW} was adopted and the kinetic energy cutoff was set to be 400 eV. The first Brillouin zone (BZ) was sampled by Monkhorst-Pack meshes method \cite{MPsampling} with a 9$\times$9$\times$1 $k$-point grid for primitive cell.
A large vacuum region with a thickness of 25 $\text\AA$ was added in the $z$ direction to avoid spurious interaction between adjacent slabs. 
Due to the strong metallicity of monolayer \ch{PtBi2}, a second order Methfessel-Paxton smearing method was used for structural relaxations and total energy calculation.
In addition, since both Pt and Bi are relatively heavy elements, spin-orbit coupling (SOC) is taken into account in the structure relaxation and self-consistent calculations.
The force and energy convergence criteria were set to  $0.01~\text{eV/\AA}$ and $10^{-6} \ \rm{eV}$, respectively. 
The dipole correction \cite{dipole_correction1,dipole_correction2} was considered during the calculations.
The phonon spectrum calculations were implemented in the PHONOPY code \cite{phonopy-phono3py-JPCM,phonopy-phono3py-JPSJ} based on the finite displacement method.
For the \textit{ab initio} molecular dynamics simulations, a $4\times4\times1$ supercell was employed to sample the NVT canonical ensemble. The system temperature was regulated by a Nosé-Hoover thermostat \cite{Nose1,Nose2,Nose3} with the Nosé mass parameter set to 0.5.

To calculate the linear and nonlinear transport response coefficients, a tight-binding Hamiltonian consisting of Pt-$5d$ and Bi-$6p$ orbitals was constructed using the Wannier90 package \cite{wannier90_2020}. 
The momentum-space integrals specified in Secs.~\ref{section:cond} through \ref{section:NLHE} were then evaluated with an in-house parallelized Julia code.
The first BZ was sampled with a dense $1000\times1000\times1$ $k$-point mesh, and the numerical convergence was carefully verified using a denser $1500\times1500\times1$ mesh.
The Dirac delta function was approximated by Gaussian function with a broadening width of 20 meV. 
The relaxation time $\tau$ was set to $20~\mathrm{fs}$, which is a moderate value compared to common metals \cite{metal_relaxtion_time} and metal oxides \cite{10.1063/1.3562141,PhysRevB.95.205202}.
To obtain 3D-equivalent response coefficients and remove the artificial volume dependence introduced by the vacuum region, an effective thickness of $L_{\mathrm{eff}} = 10~\text{\AA}$ was utilized. The conversion is given by:
\begin{equation}\label{eq:eff}
	\mathcal{O}_{\mathrm{eff}} = \frac{L_\mathrm{slab}}{L_{\mathrm{eff}}} \mathcal{O}_{\mathrm{slab}}
\end{equation}
where $L_\mathrm{slab}$ is the total lattice parameter of the simulation cell in the $z$ direction. 
With the exception of the Edelstein coefficient $\chi_{i;a}$ and the Berry curvature dipole $D_{ab}$, Eq.~(\ref{eq:eff}) is applied to all calculated response coefficients to facilitate direct comparison with traditional bulk materials.

\section{Phase transition and Transition temperature}\label{section:curie}

\begin{figure}[htbp]
	\centering
	\includegraphics[width=\linewidth]{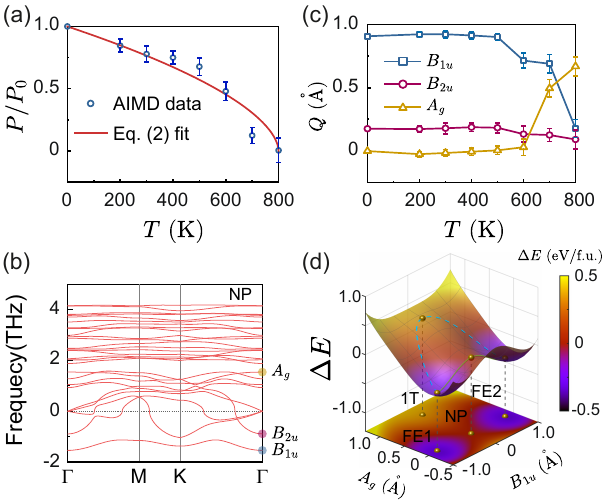}
	\caption{
		(a) Temperature-dependent normalized ferroelectric polarization $P/P_0$ obtained from AIMD simulations (blue open circles) and the corresponding critical exponent fit (red solid line) yielding a Curie temperature $T_{\mathrm{C}} \approx 800~\mathrm{K}$. 
		(b) Phonon dispersion of the NP state. The key zone-center phonon modes ($B_{1u} $, $B_{2u}$, and $A_g$) involved in structural transitions are highlighted at the $\Gamma$ point. 
		(c) Temperature evolution of the normal mode coordinates $Q$ for the $B_{1u}$, $B_{2u}$, and $A_g$ phonons extracted from equilibrium structures in the AIMD simulations. 
		(d) 3D energy landscape $\Delta E$ mapped as a function of $Q(B_{1u})$ and $Q(A_g)$. The surface illustrates the relative stabilities and transition energy barriers among the FE1, FE2, NP, and 1T states.
	}
	\label{fig:fig1} 
\end{figure}

Before investigating the electrical transport properties of the intrinsic 2D FE metal \ch{PtBi2}, it is very helpful to evaluate its Curie transition temperature to provide experimental guidance.
We employ \textit{ab initio} molecular dynamics (AIMD) simulations for the \ch{PtBi2} monolayer at different temperatures.
At each temperature, after the structures reach dynamic equilibrium, we calculate the FE polarization of every structure during the AIMD simulations. The average polarization and its standard deviation are plotted in Fig.~\ref{fig:fig1}(a).
We fit the calculated temperature-dependent polarization $P(T)$ using the following equation \cite{PhysRevB.90.014105,PhysRevLett.117.097601,newman1999monte}:
\begin{align}
	P(T) = 
	\begin{cases} 
		P_0 \left(1 - \frac{T}{T_{\mathrm C}} \right)^\delta & T < T_{\mathrm C} \\
		0 & T > T_{\mathrm C}
	\end{cases}
\end{align} 
where the $P_0$ is FE polarization at zero temperature. 
The fitting results show that it has a high Curie temperature $T_{\mathrm{C}} \approx 800~\mathrm{K}$, which ensures its room-temperature ferroelectricity. 
The fitted critical exponent $\delta$ is approximately 0.57, which is close to the mean-field approximation result of 0.5 for the 2D Ising model \cite{newman1999monte} and is higher than those of 2D FE group-IV monochalcogenides \cite{PhysRevLett.117.097601} and 3D FE \ch{PbTiO3} \cite{PhysRevB.90.014105}.

As shown in previous work \cite{PhysRevLett.133.186801}, the FE phase transition of the \ch{PtBi2} monolayer is mainly contributed by two zone-center soft modes [$B_{1u}$ and $B_{2u}$ in Fig.~\ref{fig:fig1}(b)] of the centrosymmetric nonpolar (NP) state.  
To further elucidate their role in the phase transition process, we perform the following lattice dynamics analysis \cite{dove2011introduction} and quantitative calculation.

The NP state of \ch{PtBi2} monolayer is a periodic crystal. We define the equilibrium position of the $\kappa$-th atom in the $l$-th primitive cell as $\boldsymbol{R}_{l\kappa}^0$.
The relative displacements of atoms can be written as $\boldsymbol{u}_{l\kappa} = \boldsymbol{R}_{l\kappa} - \boldsymbol{R}_{l\kappa}^0$. 
The $\boldsymbol{u}_{l\kappa}$ can be decomposed into a linear combination of phonon modes, which means \cite{Togo_2023}
\begin{align}
	u_{l\kappa\alpha} &= \sum_{\boldsymbol{q}\nu} Q(\boldsymbol{q}\nu) \frac{1}{\sqrt{N m_{\kappa}}} W_{\kappa\alpha}(\boldsymbol{q}\nu) \mathrm{e}^{i\boldsymbol{q} \cdot \boldsymbol{R}_{l\kappa}^{0}} \label{eq:displace_u} 
\end{align}
where $\alpha = x,y,z$ denotes  the Cartesian coordinates and $m_{\kappa}$ is the mass of the $\kappa$-th atom.
$\nu$ labels the phonon band index, and the composite index $\boldsymbol{q}\nu$ is used to consider a phonon mode.
$W_{\kappa\alpha}$  denotes the eigenvector of the corresponding phonon mode and $N$ is the number of the primitive cells.

When the structures of the NP state and another state are given, a set of relative atomic displacements $\boldsymbol{u}_{l\kappa}$ is determined.
The normal mode coordinate $Q(\boldsymbol{q}\nu)$ quantitatively represents the contribution of the $\boldsymbol{q}\nu$ phonon to these atomic displacements and can be obtained in the inverted transformation of Eq. (\ref{eq:displace_u}).
\begin{align}
	Q(\boldsymbol{q}\nu) = \sum_{l\kappa\alpha} \sqrt{\frac{m_{\kappa}}{N}} u_{l\kappa\alpha} W_{\kappa\alpha}^* \mathrm{e}^{-i\boldsymbol{q} \cdot \boldsymbol{R}_{l\kappa}^{0}} \label{eq:normal_Q}
\end{align} 
We calculate the $Q(\boldsymbol{q}\nu)$ of all $\Gamma$-point phonons and find that the value of $Q(B_{1u})$ is significantly higher than those of the other phonon modes. 
This result is consistent with our previous conclusion that the $B_{1u}$ is the dominant phonon mode responsible for driving the structural transition from the NP state to the FE state.

In the AIMD simulations, we also do the similar lattice dynamical analysis for the equilibrium structures at each temperature [see Fig.~\ref{fig:fig1}(c)]. 
Obviously, as the temperature increases, when all the normal mode coordinates $Q(\boldsymbol{q}\nu)$ approach zero, the crystal transitions from the FE ground state to the NP state.
The temperature-dependent $Q(B_{1u})$ and $Q(B_{2u})$ in Fig.~\ref{fig:fig1}(c) indeed confirm this trend clearly. 
However, when we carefully examine the contributions of other phonon modes during this heating process, $Q(A_g)$ exhibits a significant increase. 
In contrast to $B_{1u}$ and $B_{2u}$, the $A_g$ mode is symmetric under the spatial inversion operation $\mathcal P$ and corresponds to a transition from the NP state to the 1T state [see Fig.~\ref{fig:fig2}(a) for the corresponding structures]. 
This indicates that a ferroelectric-to-paraelectric phase transition indeed occurs around $T \approx 800~\mathrm{K}$. However, the high-temperature paraelectric state may involve competition between the NP state and the 1T state.

To clarify this issue from an energetic perspective, the energy landscapes as a function of $Q(B_{1u})$ and $Q(A_g)$ are plotted in Fig.~\ref{fig:fig1}(d). 
We find that the energy barrier of the $\mathrm{FE1} \rightarrow \mathrm{1T} \rightarrow \mathrm{FE2}$ transition path is roughly twice that of the $\mathrm{FE1} \rightarrow \mathrm{NP} \rightarrow \mathrm{FE2}$ path. Therefore, unlike the preceding dynamic analysis, a direct transition into the 1T phase does not seem energetically favored. 
Because of this, further experimental confirmation is required to pinpoint the exact structure of the high-temperature paraelectric state.
Fortunately, as we will indicate in the following section of electrical transport, simple measurements of the in-plane conductivity can be proposed as an effective experimental tool to explicitly distinguish between the 1T and NP states.

\section{Semiclassical Boltzmann transport theory and conductivity}\label{section:cond}

In a periodic crystal, the electron wave function takes the Bloch form, \ie $\psi_{n\boldsymbol{k}}(\boldsymbol{r}) = \langle \boldsymbol{r}| n\boldsymbol{k} \rangle = e^{i\boldsymbol{k}\cdot\boldsymbol{r}} \langle \boldsymbol{r}| u_{n\boldsymbol{k}} \rangle $. 
The Bloch electrons under weak electric field can be described by the following semiclassical equations of motion \cite{RevModPhys.82.1959}
\begin{align}
	\dot{\boldsymbol{r}} &= \frac{1}{\hbar} \frac{\partial \varepsilon_{n\boldsymbol{k}}}{\partial \boldsymbol{k}} - \dot{\boldsymbol{k}} \times \boldsymbol{\Omega}_n({\boldsymbol{k}}) \label{eq:Boltz_r}\\
	\hbar \dot{\boldsymbol{k}} &= -e \boldsymbol{E} 
\end{align}
where $\varepsilon_{n\boldsymbol{k}}$ is the band dispersion and $\boldsymbol{v}_{n\boldsymbol k} = \partial_{\boldsymbol k} \varepsilon_{n\boldsymbol k}/\hbar$ is the band velocity. The electron charge is taken as $-e$ (\ie $e>0$). 
The last term on the right-hand side of Eq.~(\ref{eq:Boltz_r}) is the anomalous velocity \cite{RevModPhys.82.1959, RevModPhys.82.1539, PhysRevLett.92.037204,PhysRevB.108.085120,LI2025187}, where  
$\boldsymbol{\Omega}_n({\boldsymbol{k}}) = \nabla \times \boldsymbol{A}_n({\boldsymbol{k}})$ denotes the Berry curvature with Berry connection defined as $\boldsymbol{A}_n({\boldsymbol{k}}) = i \langle u_{n\boldsymbol{k}} | \nabla_{\boldsymbol{k}} | u_{n\boldsymbol{k}} \rangle$. 
The macroscopic average of a transport quantity is computed by summing the expectation value of its associated operator over all the Bloch states, leading to the general expression 
\begin{align}\label{eq:integral}
	\langle \mathcal O \rangle =  \sum_n \int [d \boldsymbol{k}] \bra{n\boldsymbol k} \hat{\mathcal O} \ket{n\boldsymbol k} f_{n\boldsymbol k} (\boldsymbol r, t) 
\end{align}
where $\int [d \boldsymbol{k}] \equiv \int d^3 \boldsymbol{k}/(2\pi)^3$ denotes integration over the Brillouin zone, $\hat{\mathcal O}$ is the Hermitian operator corresponding to the observable and $f_{n\boldsymbol k}(\boldsymbol{r}, t)$ is the distribution function.

The distribution function $f_{n\boldsymbol k}(\boldsymbol{r}, t)$ can be determined by solving the semiclassical Boltzmann transport \cite{ashcroftsolid}
\begin{align}
	\frac{\partial f_{n\boldsymbol k}}{\partial t} + \dot{\boldsymbol{r}} \cdot \frac{\partial f_{n\boldsymbol k}}{\partial \boldsymbol{r}} + \dot{\boldsymbol{k}} \cdot \frac{\partial f_{n\boldsymbol k}}{\partial \boldsymbol{k}} = \left( \frac{\partial f_{n\boldsymbol k}}{\partial t} \right)_{\text{coll}}
\end{align}
For a homogeneous system in the steady state, where
$\partial f_{n\boldsymbol k}/\partial t$ and $\partial f_{n\boldsymbol k}/\partial \boldsymbol{r}$ vanish and the collision integral is treated within the relaxation time approximation, the Boltzmann equation simplifies to 
\begin{align}\label{eq:Boltzmann}
	\dot{\boldsymbol{k}} \cdot \frac{\partial f_{n\boldsymbol k}}{\partial \boldsymbol{k}} = -\frac{f_{n\boldsymbol k}-f_{n\boldsymbol k}^0}{\tau} 
\end{align}
where $f_{n\boldsymbol k}^{0}$ denotes the equilibrium Fermi-Dirac distribution, and $\tau$ represents the characteristic relaxation time. Solving above equation and keeping only the linear order of $\boldsymbol E$, we get
\begin{align}\label{eq:distribution}
	f_{n\boldsymbol k} = f_{n\boldsymbol k}^{0} + e\tau \boldsymbol{E} \cdot \boldsymbol{v}_{n\boldsymbol k} \frac{\partial f_{n\boldsymbol k}^0}{\partial \varepsilon_{n\boldsymbol{k}}}
\end{align}

\begin{figure}[htbp]
	\centering
	\includegraphics[width=\linewidth]{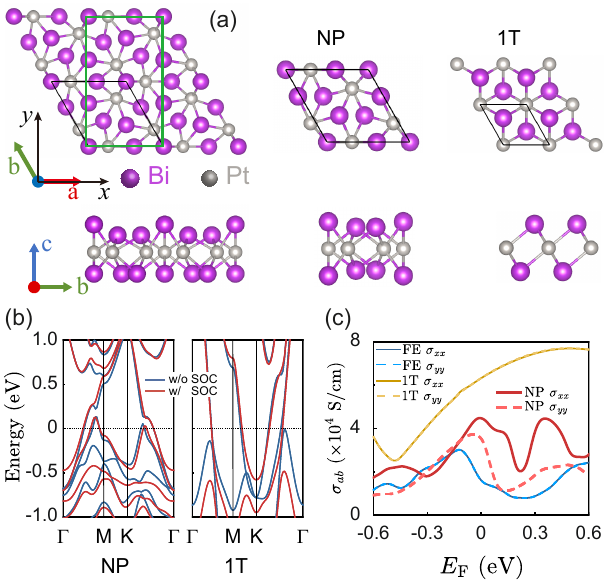}
	\caption{
		(a) Top and side views of the atomic structures for the FE (left), NP (middle), and 1T (right) states. The Bi and Pt atoms are represented by purple and grey spheres, respectively. 
		The black rhombuses mark the lattice of the pristine primitive cell, and the green rectangle marks the lattice of the FE \ch{PtBi2} under uniaxial strain, which breaks the $\mathcal{C}_{3z}$ rotational symmetry.
		(b) Calculated electronic band structures for the NP and 1T states. 
		(c) Calculated in-plane electrical conductivity tensor components ($\sigma_{xx}$ and $\sigma_{yy}$) as a function of the Fermi level. The FE and 1T states exhibit isotropic in-plane conductivity ($\sigma_{xx} = \sigma_{yy}$) protected by their $\mathcal C_{3z}$ spatial symmetry, whereas the NP state displays distinct in-plane anisotropy ($\sigma_{xx} \neq \sigma_{yy}$) due to the broken threefold rotation symmetry.
	}
	\label{fig:fig2} 
\end{figure}

The conductivity tensor is the linear response coefficient relating the electric current to an external electric field. 
Specifically, according to Eq. (\ref{eq:Boltz_r}), the electric current operator satisfies $\bra{n\boldsymbol k} \hat j^a \ket{n\boldsymbol k} = -e \dot r^a = -e (v^a_{n\boldsymbol k} + \epsilon_{abc}\Omega_{n,b}\dot{k}_c)$ with $\epsilon_{abc}$ is the Levi-Civita symbol. 
Combining Eq.~(\ref{eq:integral}) and Eq.~(\ref{eq:distribution}), the linear conductivity tensor consists of the following two parts
\begin{align}
	\sigma_{ab} &= -e^2 \tau \sum_n \int [d \boldsymbol{k}] v^a_{n\boldsymbol k}v^b_{n\boldsymbol k}  \frac{\partial f_{n\boldsymbol k}^0}{\partial \varepsilon_{n\boldsymbol{k}}} \label{eq:conductivity} \\
	\sigma_{ab}^{\mathrm{AHC}} &=  -\frac{e^2}{\hbar}  \sum_n \int [d \boldsymbol{k}] \epsilon_{abc} f_{n\boldsymbol k}^0 \Omega_{n,c}   \label{eq:AHC}
\end{align}
where the Eq.~(\ref{eq:AHC}) is the intrinsic anomalous Hall contribution arising from the anomalous velocity induced by the Berry curvature \cite{RevModPhys.82.1959, RevModPhys.82.1539, PhysRevLett.92.037204}.

From the perspective of symmetry,time-reversal symmetry $\mathcal T$ in \ch{PtBi2} forces $\sigma_{ab}^{\mathrm{AHC}}$ to vanish, so the conductivity $\sigma_{ab}$ in Eq. (\ref{eq:conductivity}) is the only term that we should consider. 
As shown in Fig.~\ref{fig:fig2}(a), the FE state and the 1T state correspond to space groups $P31m$ (No.~157) and $P\bar{3}m1$ (No.~164), respectively, both of which possess $\mathcal C_{3z}$ symmetry, resulting in isotropic in-plane conductivity, \ie  $\sigma_{xx} = \sigma_{yy}$.  
The NP state, however, breaks this threefold rotation symmetry and only possesses $\mathcal C_{2x}$ and $\mathcal M_{y}$ symmetries. 
As a result, the NP state exhibits anisotropic in-plane conductivity, \ie $\sigma_{xx} \neq \sigma_{yy}$. 
The numerical results in Fig.~\ref{fig:fig2}(c) fully reflect this difference.
As previously stated, measuring the in-plane conductivity can serve as an effective tool to determine the structure of paraelectric state.
Specifically, if the FE $\rightarrow$ NP phase transition occurs, an enhanced conductivity along with anisotropic in-plane conductivity ($\sigma_{yy}/\sigma_{xx} = 0.79$) will be experimentally observed.
If the FE $\rightarrow$ 1T phase transition occurs, the isotropic in-plane conductivity is maintained, but a significantly enhanced conductivity will be experimentally observed ($\sigma_{\mathrm{1T}}/\sigma_{\mathrm{FE}} = 3.93$).

\section{Edelstein Effect}

In addition to the electric current, the 2D FE metal \ch{PtBi2} can also exhibit an electric-field-induced magnetization, known as the Edelstein effect \cite{edelstein1990spin,Johansson_2024}. 
The magnetization induced by the linear response to the applied electric field is given by
\begin{align}\label{eq:Edelstein}
	\delta  M_i = \chi_{i;a} E_a 
\end{align}
In each unit cell, $\delta M_i$ is proportional to the spin angular momentum induced by the electric field, \ie
\begin{align}
	\delta  M_i = - \frac{\mu_{\text B}g_sV_0}{\hbar}(\langle \hat{s}^i \rangle - s_0^i)  = \chi_{i;a} E_a 
\end{align}
where $\hat{\boldsymbol s} = \hbar/2 \hat{\boldsymbol {\sigma}}$ is the spin operator and $\hat{\boldsymbol {\sigma}}$ are Pauli matrices. $\boldsymbol{s}_0$ is the spin polarization in the absence of an electric field, which is zero in \ch{PtBi2}.
$V_0$ is the volume of the unit cell, $\mu_{\text B}$ is the Bohr magneton and $g_s \approx 2$ is the electron spin $g$-factor.
Using Eq.~(\ref{eq:integral}), we obtain the Edelstein coefficient \cite{luo2026Hybridmagnetism}
\begin{align}\label{eq:Edelstein_spin}
	\chi_{i;a} = -\frac{e \tau \mu_{\text B}g_s V_0}{\hbar} \sum_n \int [d \boldsymbol{k}] s^i_{n\boldsymbol k}v^a_{n\boldsymbol k}  \frac{\partial f_{n\boldsymbol k}^0}{\partial \varepsilon_{n\boldsymbol{k}}} 
\end{align}
where $s^i_{n\boldsymbol k} = \bra{n\boldsymbol k} \hat s^i \ket{n\boldsymbol k}$ are spin expectation values.

For 2D materials, we restrict $i \in \{x, y, z\}$ and $a \in \{x, y\}$, which are physically meaningful in Eq.~(\ref{eq:Edelstein_spin}). 
According to Eq.~(\ref{eq:Edelstein}), 
the mirror operation $\mathcal M_y$ leads to the vanishing of the diagonal elements of $\chi_{i;a}$. 
The $\mathcal C_{3z}$ symmetry causes the off-diagonal elements to have opposite signs and the final result is 
\begin{align}
	\boldsymbol{\chi} = 
	\begin{bmatrix}
		0 & \chi_{x;y}  \\
		-\chi_{x;y} & 0  \\
		0 & 0 
	\end{bmatrix}
\end{align}
As discussed in previous work, these two FE states of monolayer \ch{PtBi2} are related by an inversion operator $\mathcal P$ \cite{PhysRevLett.133.186801}.
As a result, the FE1 and FE2 states exhibit Edelstein coefficients that are equal in magnitude but opposite in sign, and can be switched by a perpendicular electric field.

\begin{figure*}[htbp]
	\centering
	\includegraphics[width=\textwidth]{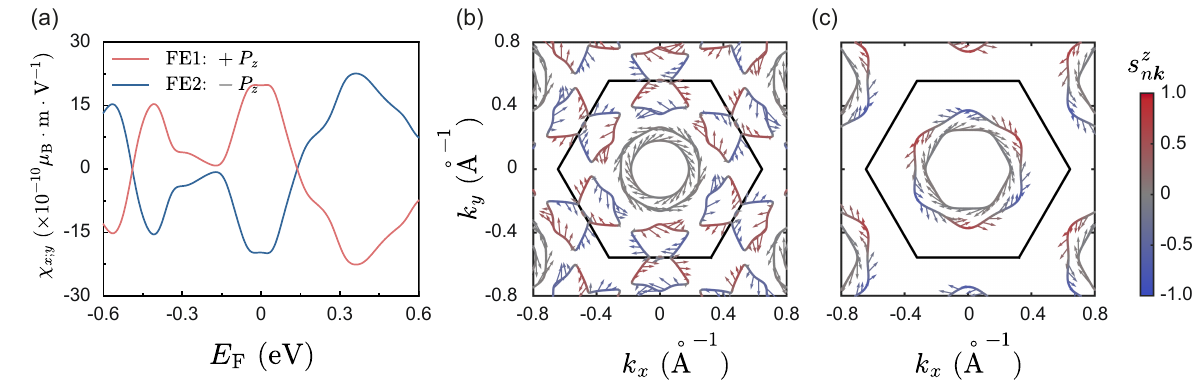}
	\caption{
	(a) The calculated Edelstein coefficient $\chi_{x;y}$ as a function of the Fermi level $E_{\text{F}}$ for FE1 ($+P_z$) and FE2 ($-P_z$) of monolayer \ch{PtBi2}. 
	(b) The Fermi surface and corresponding spin texture in the BZ at $E_{\text{F}} = 0$. 
	(c) The Fermi surface and spin texture at a shifted Fermi energy of $E_{\text{F}} = 0.2~\text{eV}$. In (b) and (c), the arrows represent the direction and magnitude of in-plane spin polarization, while the color encodes the magnitude of out-of-plane spin component $s_{n\boldsymbol{k}}^z$. 
	}
	\label{fig:fig3} 
\end{figure*}

As shown in Fig.~\ref{fig:fig3}(a), the numerical results fully agree with the above symmetry analysis. 
The factor $\partial f_{n\boldsymbol{k}}^0 / \partial \varepsilon_{n\boldsymbol{k}}$ in Eq.~(\ref{eq:Edelstein_spin}) indicates that $\chi_{i;a}$ is determined by the electronic states near the Fermi surface.
We plot the Fermi surface and its spin texture at the Fermi level $E_{\text{F}} = 0$. One can see an electron pocket with Rashba-like spin texture around the $\Gamma$ point and a hole pocket around the $\mathrm M$ point in the BZ. 
As revealed in a recent work by Pan \ea \cite{j5s5-m7j5}, the electron pocket and hole pocket contribute oppositely to $\chi_{x;y}$. When the Fermi level is shifted upward, the electron pocket expands and the hole pocket shrinks [see Fig.~\ref{fig:fig3}(c) with $E_{\text{F}} = 0.2~\text{eV}$], leading to a monotonic decrease and ultimately causing a sign reversal \cite{j5s5-m7j5}.

\section{Spin Hall Effect}
Generating spin current is a central topic in spintronics and is worth investigating in the FE metal \ch{PtBi2} as well.
The linear response of the spin current to an applied electric field, known as the spin Hall effect \cite{RevModPhys.87.1213,PhysRevLett.92.126603,PhysRevLett.94.226601,PhysRevLett.95.156601,PhysRevLett.100.096401}, is a typical approach for spin-current generation.

In analogy with the electric current operator $\hat{j}^a$ in Sec.~\ref{section:cond}, the spin current operator can be defined as
\begin{align}
	\hat j^{a,s^i} &= \frac{1}{2}(\hat v^a \hat s^i + \hat s^i \hat v^a) = e\cdot\frac{\hbar}{2e}\mathcal{J}^{a,s^i} \\
	\hat{\mathcal{J}}^{a,s^i} &= \frac{1}{2}(\hat v^a \hat\sigma^i + \hat\sigma^i \hat v^a)
\end{align}
Since the Pauli matrix $\hat\sigma^i$ is dimensionless, the operator $\hat{\mathcal{J}}^{a,s^i}$ shares the same dimension with the velocity operator $\hat v^a$. Hence, the spin current and the electric current differ merely by a factor of $\hbar / 2e$.
Based on this, we normalize
the spin Hall conductivity by the factor $\hbar / 2e$ \cite{PhysRevB.86.165108,PhysRevMaterials.4.114202,pei2025spininjection} to enable a direct comparison with the charge conductivity in Sec.~\ref{section:cond}.

Using Eq.~(\ref{eq:integral}), we obtain the spin Hall conductivity,
\begin{align} \label{eq:sigma_ext}
	\sigma_{\text{ext}}^{i;ab} = \frac{\hbar}{2e} \cdot e^2 \tau \sum_n \int [d \boldsymbol{k}] \mathcal{J}^{a,s^i}_{n\boldsymbol k}v^b_{n\boldsymbol k}  \frac{\partial f_{n\boldsymbol k}^0}{\partial \varepsilon_{n\boldsymbol{k}}}
\end{align}
where $\mathcal{J}^{a,s^i}_{n\boldsymbol k} = \bra{n\boldsymbol k} \hat{\mathcal{J}}^{a,s^i} \ket{n\boldsymbol k} $.
$\sigma_{\text{ext}}^{i;ab}$ is an odd function under the time-reversal operation $\mathcal{T}$ and vanishes in nonmagnetic materials, so some literature refers to it as the magnetic spin Hall effect (MSHE) \cite{PhysRevB.106.024410,kimata2019magnetic,PhysRevB.90.174423,PhysRevLett.104.186403}.
Although the MSHE has been intensively investigated in antiferromagnets recently \cite{luo2026Hybridmagnetism,ma2021multifunctional,PhysRevLett.126.127701,PhysRevLett.119.187204}, only the intrinsic spin Hall effect (ISHE) — arising from the quantum geometry of the electronic structure (\ie spin Berry curvature) — exists in the 2D FE metal \ch{PtBi2}.

The spin Hall conductivity of ISHE is given by 
\begin{align} \label{eq:sigma_int}
	\sigma_{\text{int}}^{i;ab} &= \frac{\hbar}{2e} \cdot \frac{e^2}{\hbar} \sum_n \int [d \boldsymbol{k}] f_{n\boldsymbol k}^0 \Omega_{n,ab}^{s^i} \\ 
	\Omega_{n,ab}^{s^i} &= -2 \operatorname{Im} \sum_{m \neq n} \frac{\mathcal{J}_{nm}^{a, s^i} v_{mn}^b}{(\omega_m - \omega_n)^2} 
\end{align}
where $\Omega_{n,ab}^{s^i}$ is the spin Berry curvature which has the similar form with the Berry curvature.
Under symmetry constraints, there are only 2 independent components and the third-order tensor $\sigma_{\text{int}}^{i;ab}$ takes the following form
\begin{align}
	\boldsymbol{\sigma}_{\text{int}}^{s^x} &= 
	\begin{bmatrix}
		0 & \sigma_{\text{int}}^{x;xy}  \\
		\sigma_{\text{int}}^{x;xy} & 0  \\ 
	\end{bmatrix} \\
	\boldsymbol{\sigma}_{\text{int}}^{s^y} &= 
	\begin{bmatrix}
		\sigma_{\text{int}}^{x;xy}  & 0  \\
		0 & -\sigma_{\text{int}}^{x;xy}  \\ 
	\end{bmatrix} \label{eq:unconventional}\\
	\boldsymbol{\sigma}_{\text{int}}^{s^z} &= 
	\begin{bmatrix}
		0 & \sigma_{\text{int}}^{z;xy}  \\
		-\sigma_{\text{int}}^{z;xy} & 0  \\ 
	\end{bmatrix} 
\end{align}
Note that the diagonal elements in Eq.~(\ref{eq:unconventional}) indicate that the induced spin current in \ch{PtBi2} can be parallel to the longitudinal charge current, which is usually referred to as the unconventional spin Hall effect \cite{PhysRevMaterials.6.045004}.

\begin{figure*}[htbp]
	\centering
	\includegraphics[width=\textwidth]{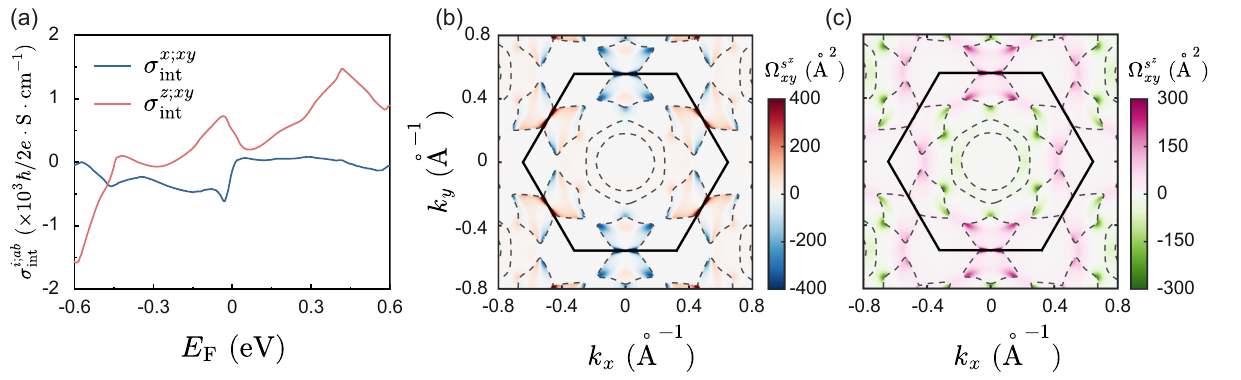}
	\caption{
	(a) The independent non-zero components of the intrinsic spin Hall conductivity, $\sigma_{\text{int}}^{x;xy}$ and $\sigma_{\text{int}}^{z;xy}$, as a function of the Fermi level $E_{\text{F}}$. 
	(b, c) The $\boldsymbol{k}$-resolved spin Berry curvature (b) $\Omega_{xy}^{s^x}$ and (c) $\Omega_{xy}^{s^z}$ summed over all occupied states in momentum space at $E_{\text{F}} = 0$. The solid black hexagon represents the first BZ, and the dashed lines indicate the Fermi surface contours. 
	}
	\label{fig:fig4} 
\end{figure*}

As shown in Fig. \ref{fig:fig4}(a), we calculate the independent components $\sigma_{\text{int}}^{x;xy}$ and $\sigma_{\text{int}}^{z;xy}$. 
Compared with most other 2D materials \cite{li2026high,zhou2025high}, \ch{PtBi2} exhibits a sizable ISHE ($\sigma_{\text{int}}^{x;xy} = -135~\hbar/2e \cdot \mathrm{S \cdot cm^{-1}}$ and $\sigma_{\text{int}}^{z;xy} = 509~\hbar/2e \cdot \mathrm{S \cdot cm^{-1}}$ at $E_{\text{F}} = 0$), rendering it promising for future applications in spintronics.
The corresponding $\boldsymbol{k}$-resolved spin Berry curvature of all occupied states is shown in Fig.~\ref{fig:fig4}(b) and (c).
We can see that at $E_{\text{F}} = 0$, the Rashba-like split band around the $\Gamma$ point provides a minor contribution, while the states near the hole pocket around the $\mathrm M$ point dominate the ISHE.

To better illustrate the charge-to-spin conversion efficiency of \ch{PtBi2}, we further calculate the spin Hall angle $\theta^{i}_{ab}$, which is a dimensionless quantity.
\begin{align} 
	\theta^{i}_{ab} &= \frac{2e}{\hbar} \frac{\abs{\sigma_{\text{int}}^{i;ab}}}{\abs{\sigma_{bb}}}
\end{align}
Combining with the electrical conductivity calculated in Sec.~\ref{section:cond}, we obtain the spin Hall angles $\theta^{x}_{xy} = 0.008$ and $\theta^{z}_{xy} = 0.032$, which are comparable to the experimental values reported for platinum ($0.056 < \theta < 0.16$) \cite{PhysRevLett.106.036601}.

\section{Nonlinear Hall Effect}\label{section:NLHE}

In previous sections, we have investigated the linear response of several typical physical quantities to an external electric field. 
The FE polarization inherently breaks the spatial inversion symmetry $\mathcal P$ in \ch{PtBi2} monolayer, and the nonlinear Hall effect (NLHE) driven by the Berry curvature dipole (BCD) \cite{PhysRevLett.115.216806,YangZhangPNAS,du2021nonlinear} is usually studied in such noncentrosymmetric systems.
The second-order nonlinear response of the electric current to the electric field can be expressed as
\begin{equation}\label{eq:NLresponse}
	j_a = \sigma_{a;bc} E_b E_c 
\end{equation}
Here, we consider an alternating electric field $E_c(t) = \text{Re}\{E_c(\omega,t)\} = \text{Re}\{\mathcal{E}_c e^{i\omega t}\}$.
Accordingly, Eq.~(\ref{eq:Boltz_r}) now becomes
\begin{align}
	\dot{r_a} 
	&= v_a - \frac{e}{2\hbar} \epsilon_{abc}\Omega_b (\mathcal{E}_c e^{i\omega t} + \mathcal{E}_c^* e^{-i\omega t})
\end{align}
For brevity, the indices $n$ and $n\boldsymbol{k}$ have been dropped in the derivation. 
Unlike the case for static electric field, under the alternating electric field the distribution function $f$ is time-dependent in the steady state, and the Boltzmann equation~(\ref{eq:Boltzmann}) becomes
\begin{align}\label{eq:Boltz}
	f-f_0 &= -\tau\frac{\partial f}{\partial t} + \frac{e\tau \boldsymbol{E}}{\hbar}\cdot \frac{\partial f}{\partial {\boldsymbol{k}}}           
\end{align} 
To solve Eq.~(\ref{eq:Boltz}), we expand the distribution function $f$ to second order in $\boldsymbol{\mathcal{E}}$:
\begin{align}\label{eq:f}
	f &= f_0 + f_1 + f_2 \notag\\
	&= f_0 + \text{Re}\{f_1^{\omega}e^{i\omega t}\} + f_2^{0\omega} + \text{Re}\{f_2^{2\omega}e^{2i\omega t} \}
\end{align}
Note that the term $f_2^{0\omega}$ accounts for the time-independent component, which originates from terms proportional to $E_b(\omega,t)E_c(-\omega,t) = \mathcal{E}_b \mathcal{E}_c^*$ \cite{boyd2020nonlinear}.

We still solve the above Boltzmann equation~(\ref{eq:Boltz}) iteratively.
Specifically, substituting Eq.~(\ref{eq:f}) into Eq.~(\ref{eq:Boltz}) and retaining terms to linear order in $\boldsymbol{\mathcal{E}}$, we obtain
\begin{align}
	f_1^{\omega} &= \frac{e\tau \mathcal{E}_a\partial_{k_a} f_0 }{(1 + i\omega\tau)\hbar} 
\end{align} 
Retaining terms to second order in $\boldsymbol{\mathcal{E}}$, we obtain \cite{PhysRevLett.115.216806,YangZhangPNAS}
\begin{align}
	f_2^{0\omega} &= \frac{(e\tau)^2 \mathcal{E}_a \mathcal{E}^*_b \partial_{k_a k_b} f_0}{2(1 + i\omega\tau)(1 - i\omega\tau)\hbar^2} \\
	f_2^{2\omega} &= \frac{(e\tau)^2 \mathcal{E}_a \mathcal{E}_b \partial_{k_a k_b} f_0}{2(1 + i\omega\tau)(1 + 2i\omega\tau)\hbar^2}
\end{align}
Similarly, the response current in Eq.~(\ref{eq:NLresponse}) can also be expanded as 
\begin{align}
	j_a &= \text{Re}\{j_a^{0\omega} + j_a^{\omega} e^{i\omega t} + j_a^{2\omega} e^{2i\omega t}\}  \notag \\
	&= -e \int [d \boldsymbol{k}] \dot{r_a} f(\boldsymbol{k})
\end{align}
Obviously, $j_a^{\omega}$ is the linear response current, $j_a^{2\omega}$ is the second-order response current, and $j_a^{0\omega}$ is the rectified second-order current.
\begin{align}
	j_a^{0\omega} &= \sigma_{a;bc}(0;\omega,-\omega){\mathcal E}_b(\omega){\mathcal E}_c(-\omega) \notag \\
	j_a^{2\omega} &= \sigma_{a;bc}(2\omega;\omega,\omega){\mathcal E}_b(\omega){\mathcal E}_c(\omega)
\end{align}
where we have use the relationship ${\mathcal E}_c(-\omega) = {\mathcal E}_c(\omega)^*$ \cite{boyd2020nonlinear}. 
Combining the above results, we finally obtain the second-order nonlinear conductivity, which consists of two contributions: a Drude contribution and a BCD contribution, given respectively by \cite{PhysRevLett.115.216806,YangZhangPNAS}
\begin{align}
	&\sigma^{\text{D}}_{a;bc}(0;\omega,-\omega) = \sigma^{\text{D}}_{a;bc}(2\omega;\omega,-\omega)  \notag \\
	&= 
	- \frac{e^3\tau^2}{2(1+i\omega\tau)(1-i\omega\tau)\hbar^2} \int [d \boldsymbol{k}] v_a \partial_{k_b k_c} f_0  \label{eq:Drude} \\
	&\sigma^{\text{BCD}}_{a;bc}(0;\omega,-\omega) =  \sigma^{\text{BCD}}_{a;bc}(2\omega;\omega,-\omega)   \notag\\ &=\frac{e^3\tau}{2(1+i\omega\tau)\hbar^2} \int [d \boldsymbol{k}] (\partial_{k_b} f_0) \epsilon_{adc} \Omega_d \label{eq:BCD} 
\end{align}
The second-order Drude term in Eq.~(\ref{eq:Drude}) is also a $\mathcal{T}$-odd function and vanishes in \ch{PtBi2}.
Using integration by parts, we can transform Eq.~(\ref{eq:BCD}) into
\begin{align}
	\sigma^{\text{BCD}}_{a;bc} 
	&=  -\epsilon_{adc} \frac{e^3\tau}{2(1+i\omega\tau)\hbar^2} \int [d \boldsymbol{k}] f_0 \partial_b \Omega_d \notag \\
	&= -\epsilon_{adc} \frac{e^3\tau}{2(1+i\omega\tau)\hbar^2} D_{bd}
\end{align}
where $D_{bd}$ is the so-called Berry curvature dipole (BCD) \cite{PhysRevLett.115.216806,YangZhangPNAS}. $D_{bd}$ is dimensionless for 3D materials and has dimensions of length for 2D materials.

\begin{figure}[htbp]
	\centering
	\includegraphics[width=\linewidth]{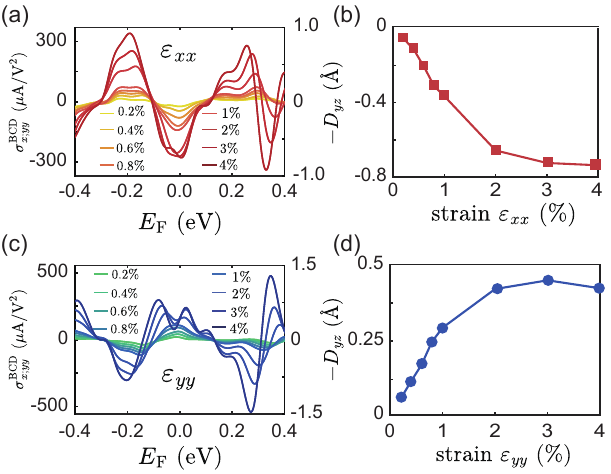}
	\caption{
	(a),(c) The nonlinear Hall conductivity $\sigma^{\text{BCD}}_{x;yy}$ and the corresponding BCD $D_{yz}$ as a function of the Fermi energy $E_{\text{F}}$ under various uniaxial strains applied along the (a) $x$ direction ($\varepsilon_{xx}$) and (c) $y$ direction ($\varepsilon_{yy}$). 
	(b),(d) The calculated BCD $D_{yz}$ at $E_{\text{F}} = 0$ as a function of the applied uniaxial strain along the (b) $x$ and (d) $y$ directions.  
	}
	\label{fig:fig5} 
\end{figure}

For a 2D system, the Berry curvature has only an out-of-plane component $\Omega_z \neq 0$.
For the FE \ch{PtBi2} monolayer, the mirror operation $\mathcal{M}_y$ forces $D_{xz} = 0$. However, the $\mathcal{C}_{3z}$ symmetry enforces $D_{yz} = D_{xz}$. Therefore, the NLHE is symmetry-forbidden.
Uniaxial strain can effectively break the $\mathcal{C}_{3z}$ rotational symmetry and induce a nonzero $D_{yz}$.
The numerical results in Fig. \ref{fig:fig5} also demonstrate that analysis.
We apply uniaxial strains ranging from 0.2\% to 4\% along the $x$ and $y$ directions in the rectangular cell [see Fig.~\ref{fig:fig2}(a)].
As shown in Fig.~\ref{fig:fig5}(b) and (d), an approximate linear increase of the BCD or the NLHE conductivity at $E_{\text{F}} = 0$ occurs for small strains ($<2\%$); beyond this, further increasing the strain does not enhance the NLHE signal.

\section{ferroelectric metal field-effect transistor}\label{section:FET}

\begin{figure}[htbp]
	\centering
	\includegraphics[width=\linewidth]{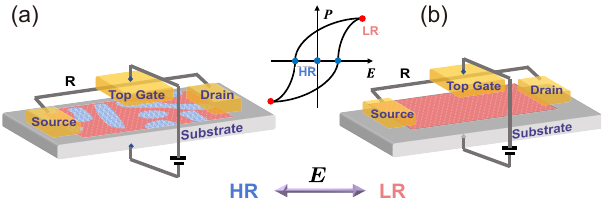}
	\caption{
	The (a) high-resistance state (HR) and (b) low-resistance state (LR) of FEM-FET. 
	}
	\label{fig:fig6} 
\end{figure}

The metal-oxide-semiconductor field-effect transistor (MOSFET) plays a central role in modern information processing systems \cite{RATNESH2021106002}.
The fundamental operating principle of the MOSFET involves the electrostatic control of a conductive channel via a gate terminal.
The resulting transition between a high-resistance (HR) state (channel pinched off) and a low-resistance (LR) state (channel turned on) allows the device to encode binary “0” and “1” data.

Unlike the MOSFET, the gate voltage does not alter the carrier concentration of the 2D FE metal, but it does change its conductivity and enables a similar nonvolatile switching between the HR and LR states.  
The underlying principle is simple and intuitive. The domain structure and the properties of domain walls have a significant impact on the transport properties of the material.
The external electric field can alter the domain
size and domain wall density of 2D FE metals, simultaneously controlling the device to switch from a LR state to a HR state (see in Fig. \ref{fig:fig6}). 
We term this device ferroelectric metal field-effect transistor (FEM-
FET), which employs a distinct mechanism to achieve
gate-controlled electric transport.

We note that the similar phenomena have been experimentally observed in ferromagnets. 
Previous studies have shown that magnetic domain walls can act as a source of resistance and can effectively modulate the electrical transport \cite{MZiese_2002,PhysRevLett.84.6090,PhysRevLett.79.5110,PhysRevB.102.054438,orfila2023large}.
Since traditional
FE materials are good insulators with high resistivity, which hinders the investigation of domain wall impacts on their electrical transport properties, the literature has shifted focus to studying the thermal transport engineered by FE domain walls \cite{liu2020bidirectional,langenberg2019ferroelectric,WANG2016220,ihlefeld2015room}. 
Obviously, the 2D FE metal, by combining ferroelectricity and metallicity, naturally circumvents these limitations and provides an opportunity to study the effect of FE domain walls on electrical transport and their applications.
Moreover, experiments have verified that \ch{PtBi2} exhibit robust superconductivity under low temperature \cite{changdar2025topological,moreno2025robust,schimmel2024surface,10.1063/10.0014014}. Upon cooling below the superconducting transition temperature, this superconductivity
may also be modulated by controlling the domain-wall
scattering with a gate voltage \cite{jindal2023coupled}.

\section{Conclusion}\label{section:conclusion}
In summary, we have systematically investigated the linear and nonlinear electrical transport properties of the intrinsic 2D FE metal \ch{PtBi2} using first-principles calculations combined with the semiclassical Boltzmann transport theory. 
Our AIMD simulations indicate a high Curie temperature reaching $800~\text{K}$, ensuring robust room-temperature ferroelectricity. 
By comparing the possible high-temperature paraelectric states, we reveal that the FE phase of \ch{PtBi2} exhibits a lower electrical conductivity than both the NP and 1T states. 
Furthermore, we propose that measuring the isotropy or anisotropy of the in-plane conductivity serves as a simple and highly effective experimental method to distinguish between the NP state and the 1T state.

Beyond the longitudinal conductivity, we have quantitatively evaluated the transverse and spin-dependent transport phenomena. The \ch{PtBi2} monolayer demonstrates a sizable Edelstein effect and an intrinsic spin Hall effect, indicating an efficient charge-to-spin conversion that holds promise for spintronic applications. 
We also demonstrate that while the NLHE is forbidden by crystal symmetry in the pristine state, a non-zero Berry curvature dipole can be induced via uniaxial strain, enabling second-order nonlinear transport. 
Finally, we propose the concept of a ferroelectric metal field-effect transistor (FEM-FET). 
By utilizing gate-voltage-modulated domain-wall scattering, this device architecture offers a novel paradigm for nonvolatile switching between high-resistance and low-resistance states. 
Overall, our comprehensive study not only sheds light on the elusive transport phenomena of intrinsic 2D FE metals but also provides solid theoretical guidance for designing next-generation spintronic and nonvolatile memory devices.
\\

\section*{Acknowledgement}
This work is supported by the National Natural Science Foundation of China (Grants No.12504107 and No. 12504216),
the Startup Project of Inner Mongolia University (Grant No. 10000-A260015/301),
the Natural Science Foundation of Hubei Province (Grant No. 2025AFB092) and the Youth Talent Project of the Science and Technology Research Program of the Education Department of Hubei Province (Grant No. Q20251221).

\bibliography{PtBi2_total.bib}

\end{document}